\documentclass[sigconf]{acmart}

\AtBeginDocument{%
  \providecommand\BibTeX{{%
    \normalfont B\kern-0.5em{\scshape i\kern-0.25em b}\kern-0.8em\TeX}}}

\setcopyright{rightsretained}
\graphicspath{ {./img/} }

\newif\ifSuppressMemo
\ifSuppressMemo
\newcommand{\memo}[1]{}
\else
\usepackage{color}
\newcommand{\memo}[1]{{\bf \textcolor{red}{[#1]}}}
\fi

\usepackage{caption}
\usepackage{subcaption}
\usepackage{siunitx}
\usepackage{amsmath}
\usepackage[acronym]{glossaries}
\glsdisablehyper
\usepackage{siunitx}
\usepackage{xcolor}
\newacronym{m2p}{M2P}{motion-to-photon}
\newacronym{3dof}{3DoF}{three degrees of freedom}
\newacronym{6dof}{6DoF}{six degrees of freedom}
\newacronym{vr}{VR}{virtual reality}
\newacronym{ar}{AR}{augmented reality}
\newacronym{mr}{MR}{mixed reality}
\newacronym{xr}{XR}{extended reality}
\newacronym{hmd}{HMD}{head-mounted display}
\newacronym{fov}{FoV}{field-of-view}
\newacronym{qoe}{QoE}{Quality of Experience}
\newacronym{sdp}{SDP}{Session Description Protocol}
\newacronym{ice}{ICE}{Interactive Connectivity Establishment}
\newacronym{ws}{WS}{WebSocket}
\newacronym{p2p}{P2P}{peer-to-peer}
\newacronym{rtt}{RTT}{round-trip time}
\newacronym{imu}{IMU}{inertial measurement unit}
\newacronym{lat}{LAT}{look-ahead time}
\newacronym{uwp}{UWP}{Universal Windows Platform}

\begin{document}
\title{Cloud Rendering-based Volumetric Video Streaming System for Mixed Reality Services}
  
\author{Serhan G{\"u}l}
\email{serhan.guel@hhi.fraunhofer.de}
\affiliation{%
  \institution{Fraunhofer HHI}}
  
\author{Dimitri Podborski}
\email{dimitri.podborski@hhi.fraunhofer.de}
\affiliation{%
  \institution{Fraunhofer HHI}}
  
\author{Jangwoo Son}
\email{jangwoo.son@hhi.fraunhofer.de}
\affiliation{%
  \institution{Fraunhofer HHI}}
  
\author{Gurdeep Singh Bhullar}
\email{gurdeepsingh.bhullar@hhi.fraunhofer.de}
\affiliation{%
  \institution{Fraunhofer HHI}}

\author{Thomas Buchholz}
\email{thomas.buchholz@telekom.de}
\affiliation{%
  \institution{Deutsche Telekom AG}}
  
\author{Thomas Schierl}
\email{thomas.schierl@hhi.fraunhofer.de}
\affiliation{%
  \institution{Fraunhofer HHI}}

\author{Cornelius Hellge}
\email{cornelius.hellge@hhi.fraunhofer.de}
\affiliation{%
  \institution{Fraunhofer HHI}}

\renewcommand{\shortauthors}{G{\"u}l and Podborski, et al.}

\copyrightyear{2020} 
\acmYear{2020} 
\acmConference[MMSys'20]{11th ACM Multimedia Systems Conference}{June 8--11, 2020}{Istanbul, Turkey}
\acmBooktitle{11th ACM Multimedia Systems Conference (MMSys'20), June 8--11, 2020, Istanbul, Turkey}\acmDOI{10.1145/3339825.3393583}
\acmISBN{978-1-4503-6845-2/20/06}

%


\begin{abstract}
Volumetric video is an emerging technology for immersive representation of 3D spaces that captures objects from all directions using multiple cameras and creates a dynamic 3D model of the scene. However, processing volumetric content requires high amounts of processing power and is still a very demanding task for today's mobile devices. To mitigate this, we propose a volumetric video streaming system that offloads the rendering to a powerful cloud/edge server and only sends the rendered 2D view to the client instead of the full volumetric content. We use 6DoF head movement prediction techniques, WebRTC protocol and hardware video encoding to ensure low-latency in different parts of the processing chain. We demonstrate our system using both a browser-based client and a Microsoft HoloLens client. Our application contains generic interfaces that allow for easy deployment of various augmented/mixed reality clients using the same server implementation.
\end{abstract}
\maketitle
\section{Introduction}
\label{sec:intro}
Recent technical advances in capturing and displaying immersive media sparked a huge market interest in \gls{vr} and \gls{ar} applications. Although the initial focus was more on omnidirectional (\ang{360}) video applications, with advances in volumetric capture technology as well as availability of mobile devices that are able to register their environment and place 3D objects at fixed places, the focus has started to shift towards volumetric video applications~\cite{schreer2019}. 

Volumetric video captures an object or scene with multiple cameras from all directions and create a dynamic 3D model of that object~\cite{schreer2019icip} . Users can view such content using an \gls{ar} device (e.g. an optical see-through \gls{ar} display) with accurate \gls{6dof} positional trackers, or on a 2D screen albeit in a less immersive fashion. Volumetric video is expected to enable novel use cases in the entertainment domain (e.g. gaming, sports replay) as well as in cultural heritage, education, and commerce~\cite{apostolopoulos2012, schreer2019icip}.

Despite the significant increases in computing power of mobile devices, rendering rich volumetric videos on such devices is still a very demanding task. Especially, presence of multiple volumetric objects in the scene can increase the processing complexity significantly. Furthermore, efficient hardware implementations for decoding of volumetric data (e.g. point clouds or meshes) are still not available, and software decoding can be very demanding in terms of battery usage and real-time rendering requirements. 

One way to reduce the processing load on the client is to send a 2D view of the volumetric object that is rendered according to the actual user position, rather then sending the entire volumetric content to the client. This is achieved by offloading the expensive rendering process to a a powerful server and transmitting the rendered views over a network to less powerful client devices. 
This technique is typically known as remote (or interactive) rendering~\cite{shi2015}. Another advantage of this approach is that network bandwidth requirements are significantly reduced because only a single 2D video is transmitted instead of full 3D volumetric content.
The rendering server can also be deployed within a cloud computing platform to provide flexible allocation of computational resources and scalability based on changes in processing load.

Recently, the reduced network latency enabled by the emerging 5G networks fostered a resurgence of interest in cloud rendering applications, especially in the domain of cloud gaming~\cite{lema2017}. Nvidia has released an SDK called CloudXR~\cite{nvidia2019} which aims to deliver advanced graphics performances to \emph{thin} clients by rendering complex immersive content on Nvidia cloud servers and streaming only the result to the clients. Google Stadia~\cite{google2019}, a cloud gaming service operated by Google, has already been launched in November 2019.

While cloud-based rendering helps to reduce the processing load on the client side, a major drawback is the added network latency which is not present in systems that perform rendering entirely on the end device. In addition to the added network latency, rendering and encoding on the server side also contribute to the increase in the end-to-end latency~\cite{guel2020}.
Degraded user experience and motion sickness are known to be common consequences of a perceivable \gls{m2p} latency~\cite{adelstein2003, allison2001}. Therefore, it is crucial to employ latency reduction techniques in every element of the processing chain.

One way to reduce latency is to serve the volumetric content from a geographically closer \emph{edge} server, thereby reducing the network latency~\cite{hu2015}. 
Also, deployment of recent real-time communication protocols such as WebRTC is crucial for meeting the demands of interactive low-latency streaming applications~\cite{holmberg2015}. 
In terms of video encoding latency, the use of fast hardware-based video encoders (e.g. Nvidia NVENC~\cite{nvenc}) is critical for reducing video compression latency while maintaining good quality and compression efficiency.
Finally, the system should predict the future position of the user using various prediction algorithms to further minimize the perceived end-to-end latency~\cite{azuma1995}.

\begin{figure*}[htbp]
    \centering
    \includegraphics[width=0.95\linewidth]{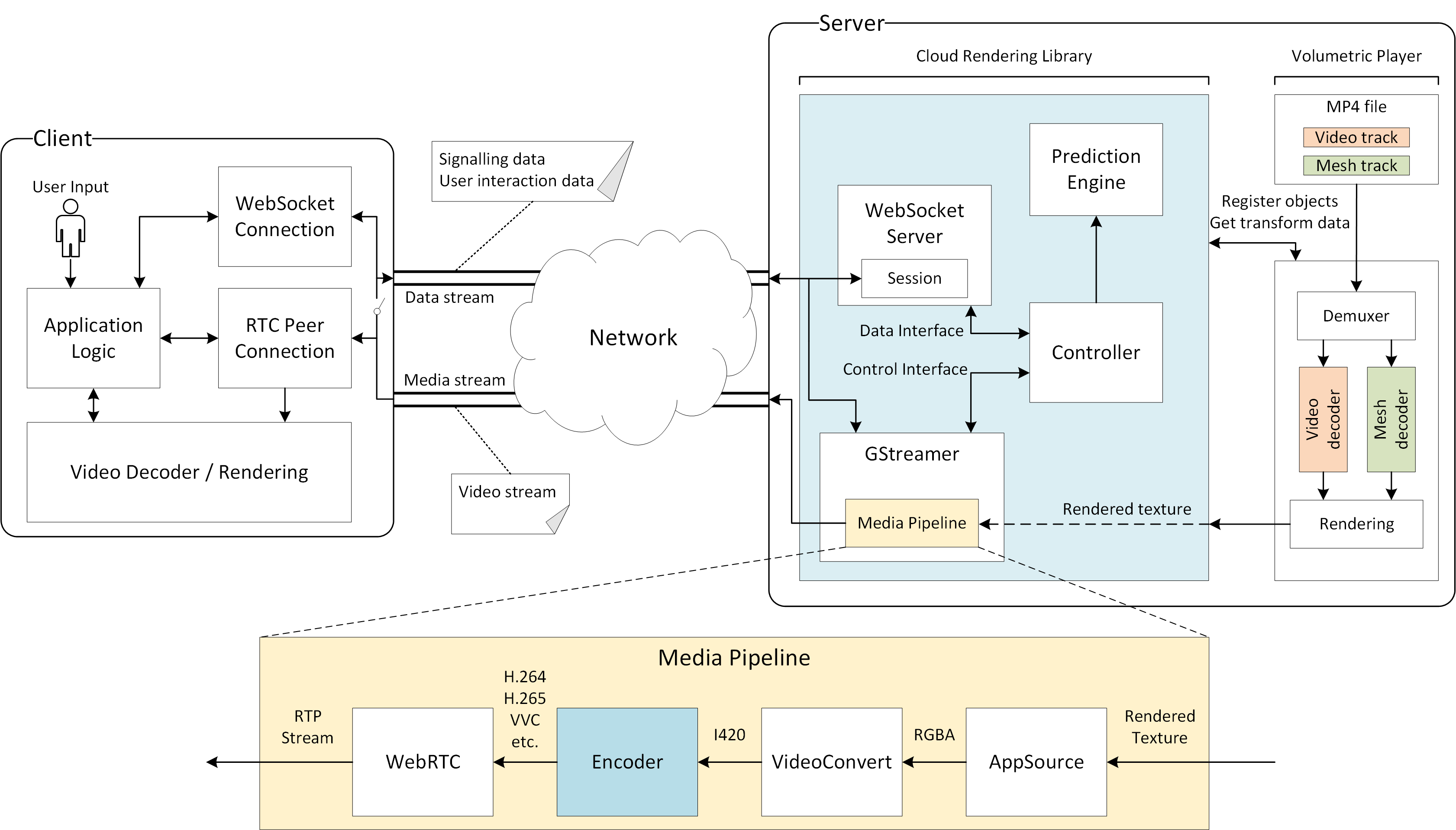}
    \caption{Overview of the system components and interfaces.}
    \Description{System architecture which is split into two parts, client and server side. On both sides WebSocket and WebRTC connections are used for communication purposes. Server side is using GStreamer pipeline which takes a rendered texture from Unity application as an input, encodes it and sends the resulting video stream to the client.}
    \label{fig:overview}
\end{figure*}

In this work, we present a system for low-latency volumetric video streaming using the cloud rendering concept that comprises a server and two different client implementations. To ensure low latency streaming, we utilize \gls{6dof} head movement prediction techniques as well as the WebRTC protocol together with Nvidia hardware encoder (NVENC). Our system provides generic interfaces such that any client implementation can be deployed using our server implementation. For our demonstration, we have implemented a client application for Microsoft HoloLens and another browser-based client that can be run on different browsers.

\section{System Architecture}
\label{sec:system}
This section describes the key components of our cloud-based volumetric video streaming system as well as the dataflow and interfaces between these components.

\subsection{Server side architecture}
\label{sec:server}
An overview of the overall system architecture is shown in Fig.~\ref{fig:overview}. Our server implementation consists of a volumetric video player and a generic cross-platform \emph{cloud rendering library} that can be integrated into different applications. 

The \textbf{volumetric video player} is implemented using Unity with several native plug-ins. The player is able to play volumetric sequences stored in a single MP4 file which consists of a video track containing the compressed texture data, and a mesh track containing the compressed mesh data. Before the start of the playback, the player registers all game objects (e.g. a volumetric object stored as an MP4 file or a virtual camera) that are to be controlled by the cloud rendering library and/or the client. After registration, the player can start playing the MP4 file by demultiplexing it and feeding the elementary streams to the corresponding video, audio and mesh decoders.
Then, each mesh is synchronized with the corresponding texture and rendered to the scene. Then, the camera representing the client's viewport captures the rendered view of the volumetric object and passes the RenderTexture to the cloud rendering library for further processing. The player concurrently asks the library for the latest positions of the previously registered game objects and updates the rendered view accordingly.

The \textbf{cloud rendering library} is a cross-platform library written in C++ that can be easily integrated into different applications. In our Unity application, we integrated it as a native plug-in into the player. The library contains various modules for application control, media processing, and the communication interfaces between the server and the client. The main modules of our library are the \Gls{ws} Server, GStreamer module, Controller, ObjectPool and Prediction Engine. Each module runs asynchronously in its own thread to achieve high performance.

The \textbf{WebSocket Server} is used for exchanging signaling data between the client and the server. Such signaling data includes \gls{sdp}, \gls{ice} as well as application-specific metadata for scene description. The \gls{ws} connection is also used for transmission of the control data in order to modify the position and orientation of any registered game object or camera. Our system also allows the usage of WebRTC data channels for control data exchange after a peer-to-peer connection is established. Both plain and secure WebSockets are supported which is important for running the system in real use cases.

The \textbf{Gstreamer} module contains the media processing pipeline which takes the rendered texture (at 60 fps) as input, compresses it as a video stream, and transmits it to the client using WebRTC. Specifically, the Unity RenderTexture is inserted into the pipeline using the \emph{appsrc} element of GStreamer.
Since the texture is in RGBA format, it has to be passed through a \emph{videoconvert} element to bring it to the I420 format accepted by the encoder. 
We use the Nvidia encoder (NVENC) to compress the texture using H.264/AVC (high profile, level 3.1, IPPP.. GOP structure, no B frames) but it is also possible to encode in H.265/HEVC using NVENC, or replace the encoder block with a different encoder e.g a software encoder such as x264.
Finally, the resulting compressed bitstream is packaged into RTP packets, encrypted, and sent to the client using the WebRTC protocol. 
For our volumetric sequence, the bitrate of the encoded bitstream (at a resolution of $1280\times720$) varies between 3-9 Mbps depending on the size of the volumetric object inside the viewport and user movement. Our system can also generate videos at 1080p and 4K resolution.

The \textbf{ObjectPool} is a logical structure that maintains the states of all registered objects (such as position, orientation and scale) enabling the client to position virtual objects correctly.

The \textbf{Controller} contains the application logic and controls the other modules depending on the application state. For example, it closes the media pipeline if a client disconnects, and re-initializes the pipeline when a new client is connected. 
The controller is also responsible for creating response messages for the client. For example, when the client requests the scene description, the WS Server notifies the controller that the request has arrived, and the controller creates a response message in JSON format with all available objects from the pool. The response message goes back to the WS Server which deals with the transmission it back to client.

The \textbf{Prediction Engine} implements a \gls{6dof} user movement prediction algorithm to predict the future head position of the user based on her past movement patterns. Based on the client interaction and the predictions output by the Prediction Engine, the Controller updates the positions of the registered objects accordingly such that the rendering engine renders a scene matching to the predicted user position that will be attained after a predefined prediction interval. The prediction interval is set to be equal to the estimated \gls{m2p} latency of the system.
Currently, an autoregressive model (as described in~\cite{guel2020}) is implemented but the module allows integration of different kind of prediction techniques e.g. Kalman filtering~\cite{azuma1995}.

\subsection{Client side architecture}
\label{sec:client}

Our client-side architecture is depicted on the left side of the Fig.~\ref{fig:overview} and essentially consists of the following modules: \acrlong{ws} Connection, WebRTC Connection, Video Decoder, Application Logic, and the client application: a browser client or a native client for Microsoft HoloLens. 

Before the streaming session starts, the client establishes a \gls{ws} connection to the server and asks the server to send a description of the rendered scene. The server responds with a list of objects and parameters. After receiving the scene description, the client replicates the scene and initiates a peer-to-peer WebRTC connection (RTCPeerConnection) with the server. The server and the client begin the WebRTC negotiation process while sending SDP and ICE data over the previously established \gls{ws} connection. Then, the peer-to-peer connection is established and the client receives a video stream over RTCPeerConnection that contains the rendered view of the 3D scene corresponding to the (predicted) user pose. The client can use the \gls{ws} connection and send control data back to the server to modify the rendered view. For example, the client may signal the changes in user's pose, or it may rotate, move or scale any volumetric object in the scene based on the user interaction.

We implemented both, a web browser player and a native application for the Microsoft HoloLens. While our browser application targets use cases in which the volumetric content is viewed on a 2D screen such as a tablet or computer display, our HoloLens client targets AR/MR applications that potentially offer better immersion and more natural interactivity.

\subsubsection*{\textbf{Browser client}}
Our browser client is implemented in JavaScript and it uses the \emph{three.js}~\cite{threejs} library that allows to interact with the volumetric object using a mouse, keyboard or touchscreen. Specifically, the user can move the camera around, drag the object to change its position and use range sliders to change the orientation or scaling of the object. The client application has been tested on different web browsers such as Chrome, Firefox, and Safari.

\subsubsection*{\textbf{Hololens client}}
Our HoloLens application is built as \gls{uwp} application with Unity. We build separate applications both for x86 and ARM architectures since HoloLens 1 executes applications on an x86 processor whereas HoloLens 2 uses an ARM processor. On the server side, the renderer designates the user as a camera in Unity 3D space and moves the camera according to the user's position. 
Due to the properties of the optical see-through display used in HoloLens, black pixels are seen fully transparent while white pixels are seen as increasingly opaque~\cite{MRrendering}. This display property can be exploited to remove the background of the volumetric object inside the video stream such that the volumetric object is perceived to be overlaid onto the real world. To achieve this, the background pixels (already rendered with a solid color on the server) are set to black in HoloLens using a shader.

The HoloLens app renders the 2D scene onto a plane orthogonal to the user's point of view in the world space (see Fig.~\ref{fig:holo_demo}). When the user changes her position and a new view is rendered on the server, this plane is always rotated towards the user. Thereby, the user perceives the different 2D views rendered onto the orthogonal plane as though a 3D object were present in the scene.

\section{Demonstration}
\label{sec:demo}
We will demonstrate our cloud-based volumetric video streaming system on the web browser client as well as the HoloLens client. In reality, we are able to deploy our server application at a 5G edge server as well as on a AWS instance. However, due to possible limitations in Internet connectivity, we will use a router to connect both clients over WiFi to a LAN.

Users will be able to interact with the volumetric content using both the browser and HoloLens clients. In both cases, users may scale, rotate, and move object.
While using the browser client, the users will also be able to measure the \gls{m2p} latency using our custom-built latency measurement tool~\cite{guel2020}.
This tool sends predefined textures (known by the client) depending on the control data received from the client. As soon as the client detects the corresponding texture based on its previous input to the server, it stops the timer and computes the \gls{m2p} latency. Running the server on an AWS instance in Frankfurt and with the WiFi connected client in Berlin we are measuring an average network latency of \SI{13.3}{ms} and a \gls{m2p} latency around \SI{60}{ms}.

\begin{figure}[htbp]
    \centering
    \includegraphics[width=0.95\linewidth]{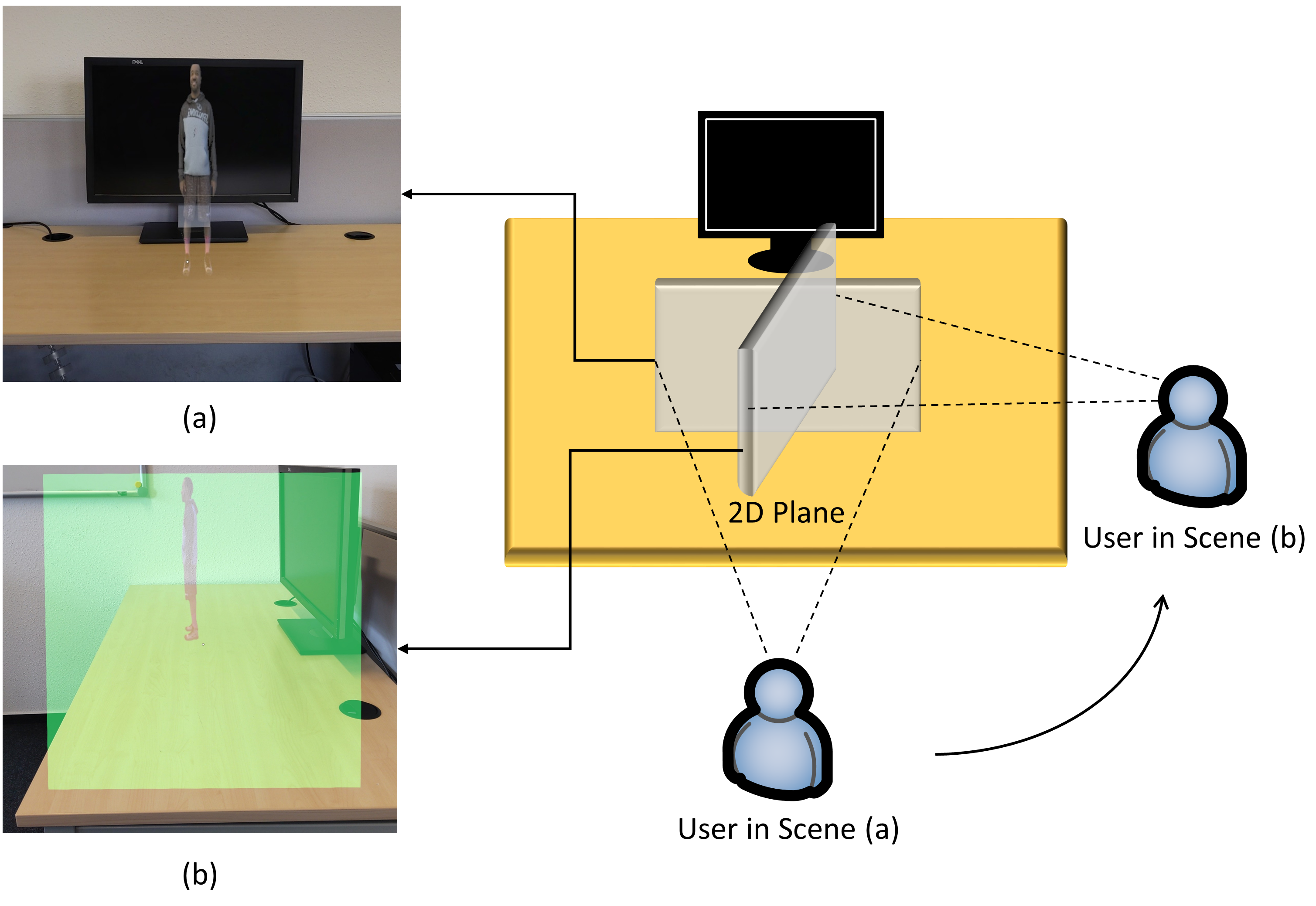}
    \caption{Captured scene from our HoloLens client; (a) Final scene after background removal (b) Scene before background removal.}
    \label{fig:holo_demo}
\end{figure}

Fig.~\ref{fig:holo_demo} shows the implementation of our HoloLens client where the transmitted 2D scene is rendered on the orthogonal plane as shown in the figure.
After receiving and decoding the 2D video, the renderer detects a certain range of RGB values in the fragment shader and adjusts the plane background in Fig.~\ref{fig:holo_demo}(b) to the transparent background as in Fig.~\ref{fig:holo_demo}(a). 
The user can then move the object to any desired position using spatial gestures designed for scaling and dragging 3D objects in HoloLens.

As supplementary material, we provide a video describing our system architecture and showing the functionality of our web browser and HoloLens clients.

\section{Conclusion}
In this paper, we presented a low-latency cloud rendering-based volumetric streaming system. Our system utilizes a powerful server for rendering of volumetric videos and decreases processing requirements and battery usage in end devices. Also, by avoiding the streaming of full 3D data, required bandwidth is significantly reduced. To reduce the increased motion-to-photon latency due to the added network latency, we use 6DoF movement prediction techniques, low latency streaming protocols (WebRTC) and low latency hardware video encoders (NVENC).
Based on the developed volumetric streaming system, our future work will include investigation of the effect of latency on the user perception in AR environments as well as development of more effective 6DoF prediction techniques.

\bibliographystyle{ACM-Reference-Format}
\bibliography{mmsys2020-demo}


\begin{thebibliography}{16}


\ifx \showCODEN    \undefined \def \showCODEN     #1{\unskip}     \fi
\ifx \showDOI      \undefined \def \showDOI       #1{#1}\fi
\ifx \showISBNx    \undefined \def \showISBNx     #1{\unskip}     \fi
\ifx \showISBNxiii \undefined \def \showISBNxiii  #1{\unskip}     \fi
\ifx \showISSN     \undefined \def \showISSN      #1{\unskip}     \fi
\ifx \showLCCN     \undefined \def \showLCCN      #1{\unskip}     \fi
\ifx \shownote     \undefined \def \shownote      #1{#1}          \fi
\ifx \showarticletitle \undefined \def \showarticletitle #1{#1}   \fi
\ifx \showURL      \undefined \def \showURL       {\relax}        \fi
\providecommand\bibfield[2]{#2}
\providecommand\bibinfo[2]{#2}
\providecommand\natexlab[1]{#1}
\providecommand\showeprint[2][]{arXiv:#2}

\bibitem[\protect\citeauthoryear{Adelstein, Lee, and Ellis}{Adelstein
  et~al\mbox{.}}{2003}]%
        {adelstein2003}
\bibfield{author}{\bibinfo{person}{Bernard~D. Adelstein},
  \bibinfo{person}{Thomas~G. Lee}, {and} \bibinfo{person}{Stephen~R. Ellis}.}
  \bibinfo{year}{2003}\natexlab{}.
\newblock \showarticletitle{Head Tracking Latency in Virtual Environments:
  Psychophysics and a Model}.
\newblock \bibinfo{journal}{\emph{Proceedings of the Human Factors and
  Ergonomics Society Annual Meeting}} \bibinfo{volume}{47},
  \bibinfo{number}{20} (\bibinfo{date}{Oct.} \bibinfo{year}{2003}),
  \bibinfo{pages}{2083--2087}.
\newblock
\urldef\tempurl%
\url{https://doi.org/10.1177/154193120304702001}
\showDOI{\tempurl}


\bibitem[\protect\citeauthoryear{Allison, Harris, Jenkin, Jasiobedzka, and
  Zacher}{Allison et~al\mbox{.}}{2001}]%
        {allison2001}
\bibfield{author}{\bibinfo{person}{R.S. Allison}, \bibinfo{person}{L.R.
  Harris}, \bibinfo{person}{M. Jenkin}, \bibinfo{person}{U. Jasiobedzka}, {and}
  \bibinfo{person}{J.E. Zacher}.} \bibinfo{year}{2001}\natexlab{}.
\newblock \showarticletitle{Tolerance of temporal delay in virtual
  environments}. In \bibinfo{booktitle}{\emph{Proceedings {IEEE} Virtual
  Reality 2001}}. \bibinfo{publisher}{{IEEE} Comput. Soc},
  \bibinfo{pages}{247--254}.
\newblock
\urldef\tempurl%
\url{https://doi.org/10.1109/vr.2001.913793}
\showDOI{\tempurl}


\bibitem[\protect\citeauthoryear{Apostolopoulos, Chou, Culbertson, Kalker,
  Trott, and Wee}{Apostolopoulos et~al\mbox{.}}{2012}]%
        {apostolopoulos2012}
\bibfield{author}{\bibinfo{person}{John~G Apostolopoulos},
  \bibinfo{person}{Philip~A Chou}, \bibinfo{person}{Bruce Culbertson},
  \bibinfo{person}{Ton Kalker}, \bibinfo{person}{Mitchell~D Trott}, {and}
  \bibinfo{person}{Susie Wee}.} \bibinfo{year}{2012}\natexlab{}.
\newblock \showarticletitle{The road to immersive communication}.
\newblock \bibinfo{journal}{\emph{Proc. IEEE}} \bibinfo{volume}{100},
  \bibinfo{number}{4} (\bibinfo{year}{2012}), \bibinfo{pages}{974--990}.
\newblock


\bibitem[\protect\citeauthoryear{Azuma}{Azuma}{1995}]%
        {azuma1995}
\bibfield{author}{\bibinfo{person}{Ronald~Tadao Azuma}.}
  \bibinfo{year}{1995}\natexlab{}.
\newblock \emph{\bibinfo{title}{Predictive tracking for augmented reality}}.
\newblock \bibinfo{thesistype}{Ph.D. Dissertation}. \bibinfo{school}{University
  of North Carolina at Chapel Hill}.
\newblock


\bibitem[\protect\citeauthoryear{Google}{Google}{2019}]%
        {google2019}
\bibfield{author}{\bibinfo{person}{Google}.} \bibinfo{year}{2019}\natexlab{}.
\newblock \bibinfo{booktitle}{\emph{Google Stadia}}.
\newblock Google Inc.
\newblock
\urldef\tempurl%
\url{https://stadia.google.com/}
\showURL{%
\tempurl}


\bibitem[\protect\citeauthoryear{G\"{u}l, Podborski, Buchholz, Schierl, and
  Hellge}{G\"{u}l et~al\mbox{.}}{2020}]%
        {guel2020}
\bibfield{author}{\bibinfo{person}{Serhan G\"{u}l}, \bibinfo{person}{Dimitri
  Podborski}, \bibinfo{person}{Thomas Buchholz}, \bibinfo{person}{Thomas
  Schierl}, {and} \bibinfo{person}{Cornelius Hellge}.}
  \bibinfo{year}{2020}\natexlab{}.
\newblock \bibinfo{title}{Low Latency Volumetric Video Edge Cloud Streaming}.
\newblock
\newblock
\showeprint[arxiv]{2001.06466v1}


\bibitem[\protect\citeauthoryear{Holmberg, Hakansson, and Eriksson}{Holmberg
  et~al\mbox{.}}{2015}]%
        {holmberg2015}
\bibfield{author}{\bibinfo{person}{Christer Holmberg}, \bibinfo{person}{Stefan
  Hakansson}, {and} \bibinfo{person}{G Eriksson}.}
  \bibinfo{year}{2015}\natexlab{}.
\newblock \showarticletitle{Web real-time communication use cases and
  requirements}.
\newblock \bibinfo{journal}{\emph{RFC 7478}} (\bibinfo{year}{2015}).
\newblock


\bibitem[\protect\citeauthoryear{Hu, Patel, Sabella, Sprecher, and Young}{Hu
  et~al\mbox{.}}{2015}]%
        {hu2015}
\bibfield{author}{\bibinfo{person}{Yun~Chao Hu}, \bibinfo{person}{Milan Patel},
  \bibinfo{person}{Dario Sabella}, \bibinfo{person}{Nurit Sprecher}, {and}
  \bibinfo{person}{Valerie Young}.} \bibinfo{year}{2015}\natexlab{}.
\newblock \showarticletitle{Mobile edge computing—A key technology towards
  5G}.
\newblock \bibinfo{journal}{\emph{ETSI white paper}} \bibinfo{volume}{11},
  \bibinfo{number}{11} (\bibinfo{year}{2015}), \bibinfo{pages}{1--16}.
\newblock


\bibitem[\protect\citeauthoryear{Lema, Laya, Mahmoodi, Cuevas, Sachs,
  Markendahl, and Dohler}{Lema et~al\mbox{.}}{2017}]%
        {lema2017}
\bibfield{author}{\bibinfo{person}{Maria~A Lema}, \bibinfo{person}{Andres
  Laya}, \bibinfo{person}{Toktam Mahmoodi}, \bibinfo{person}{Maria Cuevas},
  \bibinfo{person}{Joachim Sachs}, \bibinfo{person}{Jan Markendahl}, {and}
  \bibinfo{person}{Mischa Dohler}.} \bibinfo{year}{2017}\natexlab{}.
\newblock \showarticletitle{Business case and technology analysis for 5G low
  latency applications}.
\newblock \bibinfo{journal}{\emph{IEEE Access}}  \bibinfo{volume}{5}
  (\bibinfo{year}{2017}), \bibinfo{pages}{5917--5935}.
\newblock


\bibitem[\protect\citeauthoryear{Microsoft}{Microsoft}{2020}]%
        {MRrendering}
\bibfield{author}{\bibinfo{person}{Microsoft}.}
  \bibinfo{year}{2020}\natexlab{}.
\newblock \bibinfo{booktitle}{\emph{Mixed Reality Rendering}}.
\newblock Microsoft Corporation.
\newblock
\urldef\tempurl%
\url{https://docs.microsoft.com/en-us/windows/mixed-reality/rendering}
\showURL{%
Retrieved February 25, 2020 from \tempurl}


\bibitem[\protect\citeauthoryear{Nvidia}{Nvidia}{2019}]%
        {nvidia2019}
\bibfield{author}{\bibinfo{person}{Nvidia}.} \bibinfo{year}{2019}\natexlab{}.
\newblock \bibinfo{booktitle}{\emph{NVIDIA CloudXR Delivers Low-Latency AR/VR
  Streaming Over 5G Networks to Any Device}}.
\newblock Nvidia Corporation.
\newblock
\urldef\tempurl%
\url{https://blogs.nvidia.com/blog/2019/10/22/nvidia-cloudxr}
\showURL{%
\tempurl}


\bibitem[\protect\citeauthoryear{Nvidia}{Nvidia}{2020}]%
        {nvenc}
\bibfield{author}{\bibinfo{person}{Nvidia}.} \bibinfo{year}{2020}\natexlab{}.
\newblock \bibinfo{booktitle}{\emph{Nvidia Video Codec SDK}}.
\newblock Nvidia Corporation.
\newblock
\urldef\tempurl%
\url{https://developer.nvidia.com/nvidia-video-codec-sdk}
\showURL{%
Retrieved February 25, 2020 from \tempurl}


\bibitem[\protect\citeauthoryear{Schreer, Feldmann, Kauff, Eisert, Tatzelt,
  Hellge, M{\"u}ller, Ebner, and Bliedung}{Schreer et~al\mbox{.}}{2019a}]%
        {schreer2019}
\bibfield{author}{\bibinfo{person}{O Schreer}, \bibinfo{person}{I Feldmann},
  \bibinfo{person}{P Kauff}, \bibinfo{person}{P Eisert}, \bibinfo{person}{D
  Tatzelt}, \bibinfo{person}{C Hellge}, \bibinfo{person}{K M{\"u}ller},
  \bibinfo{person}{T Ebner}, {and} \bibinfo{person}{S Bliedung}.}
  \bibinfo{year}{2019}\natexlab{a}.
\newblock \showarticletitle{Lessons learnt during one year of commercial
  volumetric video production}. In \bibinfo{booktitle}{\emph{2019 IBC
  conference}}. IBC, \bibinfo{publisher}{IBC}.
\newblock


\bibitem[\protect\citeauthoryear{Schreer, Feldmann, Renault, Zepp, Worchel,
  Eisert, and Kauff}{Schreer et~al\mbox{.}}{2019b}]%
        {schreer2019icip}
\bibfield{author}{\bibinfo{person}{Oliver Schreer}, \bibinfo{person}{Ingo
  Feldmann}, \bibinfo{person}{Sylvain Renault}, \bibinfo{person}{Marcus Zepp},
  \bibinfo{person}{Markus Worchel}, \bibinfo{person}{Peter Eisert}, {and}
  \bibinfo{person}{Peter Kauff}.} \bibinfo{year}{2019}\natexlab{b}.
\newblock \showarticletitle{Capture and 3D Video Processing of Volumetric
  Video}. In \bibinfo{booktitle}{\emph{2019 IEEE International Conference on
  Image Processing (ICIP)}}. IEEE, \bibinfo{publisher}{IEEE},
  \bibinfo{pages}{4310--4314}.
\newblock


\bibitem[\protect\citeauthoryear{Shi and Hsu}{Shi and Hsu}{2015}]%
        {shi2015}
\bibfield{author}{\bibinfo{person}{Shu Shi} {and} \bibinfo{person}{Cheng-Hsin
  Hsu}.} \bibinfo{year}{2015}\natexlab{}.
\newblock \showarticletitle{A {Survey} of {Interactive} {Remote} {Rendering}
  {Systems}}.
\newblock \bibinfo{journal}{\emph{Comput. Surveys}} \bibinfo{volume}{47},
  \bibinfo{number}{4} (\bibinfo{date}{May} \bibinfo{year}{2015}),
  \bibinfo{pages}{1--29}.
\newblock
\showISSN{03600300}
\urldef\tempurl%
\url{https://doi.org/10.1145/2719921}
\showDOI{\tempurl}


\bibitem[\protect\citeauthoryear{three.js}{three.js}{2019}]%
        {threejs}
three.js \bibinfo{year}{2019}\natexlab{}.
\newblock \bibinfo{booktitle}{\emph{Three.js, JavaScript 3D Library}}.
\newblock
\urldef\tempurl%
\url{https://threejs.org/}
\showURL{%
Retrieved February 25, 2020 from \tempurl}
\newblock
\shownote{Version r101.}


\end{thebibliography}

\end{document}